\documentclass[12pt]{article}
\usepackage{amsthm}
\usepackage{amssymb}
\usepackage{amsmath}
\usepackage{amsfonts}
\usepackage{enumerate}
\usepackage{verbatim}
\usepackage[dvips]{graphicx}
\usepackage[english]{babel}
\usepackage[cp1250]{inputenc}
\usepackage[T1]{fontenc}
\usepackage[a4paper,left=2.8cm,right=2.8cm,top=2.8cm,bottom=2.8cm]{geometry}
\usepackage{natbib}

\newtheorem{thm}{Theorem}[section]

\newtheorem{df}{Definition}[section]

\newcommand{\Proof}{\noindent\emph{{Proof:}}\newline}
\numberwithin{equation}{section}

\newcommand{\cbdu}{\quad\hfill\mbox{$\Box$}\\}
\newcommand{\beqo}{\begin{eqnarray*}}
\newcommand{\eeqo}{\end{eqnarray*}\noindent}

\newcommand{\beq}{\begin{eqnarray}}
\newcommand{\eeq}{\end{eqnarray}\noindent}
\def\be{{\mathbb{E}}}
\def\bp{{\mathbb{P}}}
\def\bq{{\mathbb{Q}}}
\def\br{{\mathbb{R}}}
\def\bfi{{\mathcal{F}}}
\def\sj{\sum_{i=1}^{I}}
\def\var{|\sigma(t)|^2+\sum_{i=1}^I|\gamma_i(t)|^2\lambda_i(t)}

\begin{document}

\title{Instantaneous mean-variance hedging and instantaneous Sharpe ratio pricing in a regime-switching financial model, with applications to equity-linked claims}

\author{\textbf{{\L}ukasz Delong\footnote{Warsaw School of Economics, Al. Niepodleglosci 162, Warsaw, Poland} \ and Antoon Pelsser\footnote{Maastricht University, Tongersestraat 53, Maastricht, The Netherlands}}
}

\maketitle

\newpage

\begin{abstract}
\noindent We study hedging and pricing of unattainable contingent claims in a non-Markovian regime-switching financial model. Our financial market consists of a bank account and a risky asset whose dynamics are driven by a Brownian motion and a multivariate counting process with stochastic intensities. The interest rate, drift, volatility and intensities fluctuate over time and, in particular, they depend on the state (regime) of the economy which is modelled by the multivariate counting process. Hence, we can allow for stressed market conditions. We assume that the trajectory of the risky asset is continuous between the transition times for the states of the economy and that the value of the risky asset jumps at the time of the transition. We find the hedging strategy which minimizes the instantaneous mean-variance risk of the hedger's surplus and we set the price so that the instantaneous Sharpe ratio of the hedger's surplus equals a predefined target. We use Backward Stochastic Differential Equations. Interestingly, the instantaneous mean-variance hedging and instantaneous Sharpe ratio pricing can be related to no-good-deal pricing and robust pricing and hedging under model ambiguity. We discuss key properties of the optimal price and the optimal hedging strategy. We also use our results to price and hedge mortality-contingent claims with financial components (equity-linked insurance claims) in a combined insurance and regime-switching financial model.

\noindent Keywords: Counting process, instantaneous mean-variance risk, instantaneous Sharpe ratio, no-good-deal pricing, model ambiguity.

\end{abstract}

\newpage

\section{Introduction}

Pricing and hedging in incomplete markets is the most important subject in the financial and actuarial literature. Despite numerous papers, there is still a need to develop new pricing and hedging methods and derive prices and hedging strategies in realistic financial and actuarial models. In this paper we focus on instantaneous mean-variance hedging and instantaneous Sharpe ratio pricing of claims in a regime-switching financial model.

Empirical studies show that regime-switching models can explain empirical behaviors of many economic and financial data, especially the long term behavior of these data, see \citet{hamilton}, \citet{hardy}, \citet{eliotbook}. The rationale behind the regime-switching
framework is that the financial market may switch between a low-volatility
state and a high-volatility state, or even between more states representing the conditions of the economy. The switching behavior for the
states of the financial market can be attributed to structural changes in economic conditions and
changes in business environments. It is clear that there are significant fluctuations in economic variables
over a long period of time. Hence, the switching behavior of
the states of the financial market (the states of the economy) should be particulary incorporated in models used for valuation of long-term derivatives. We point out that the use of regime-switching models has been recommended by the American Academy of Actuaries and the Canadian
Institute of Actuaries for valuation of long-term financial guarantees embedded in insurance contracts.

In this paper we consider a general non-Markovian regime-switching financial model. The dynamics of a bank account and a risky asset are driven by a Brownian motion and a multivariate counting process with stochastic intensities. The interest rate, drift, volatility and intensities fluctuate over time and, in particular, they depend on the state (regime) of the economy which is modelled by the multivariate counting process. We assume that the trajectory of the risky asset is continuous between the transition times for the states of the economy and that the value of the risky asset jumps at the transition time. Since we use stochastic transition intensities, we incorporate the important feedback effect in which not only the stock price is affected by the transitions between the states of the economy but also the stock price determines the transition intensities, see \citet{elliot}. The goal is to price and hedge unattainable contingent claims in our regime-switching financial model.

Pricing and hedging in regime-switching models have gained a lot of interest in the literature, see \cite{donely2}, \cite{elliot}, \cite{elliotannals}, \citet{elliotstochastic}, \citet{siu}, \citet{siuime}, \citet{wu}, where risk minimization, quadratic hedging, multi-period Markowitz optimization, the Esscher transform are applied. We use a different pricing and hedging objective and we investigate instantaneous mean-variance hedging and instantaneous Sharpe ratio pricing. Since the optimal hedging strategy and the optimal price derived under instantaneous mean-variance hedging and instantaneous Sharpe ratio pricing can also be interpreted in terms of no-good-deal pricing and robust pricing and hedging under model ambiguity, we believe that the approach considered in this paper can be very useful for real financial applications. We should point out that in this paper we in fact investigate three pricing and hedging approaches: instantaneous mean-variance hedging and instantaneous Sharpe ratio pricing, no-good-deal pricing and pricing and hedging under model ambiguity, which are equivalent.

\citet{youngfin}, \citet{younglife}, \citet{youngannuity} were the first to apply instantaneous variance hedging and instantaneous Sharpe ratio pricing. They find the hedging strategy which minimizes the instantaneous variance (the quadratic variation) of the surplus (the difference between the hedging portfolio and the price of a claim) and set the price so that the instantaneous Sharpe ratio of the surplus equals a predefined target. \cite{youngfin} use this approach to price and hedge claims contingent on a non-tradeable financial risk, and \citet{youngannuity}, \citet{younglife} use this approach to price stochastic mortality risk in insurance models. Interestingly, the authors show the equivalence between the local variance minimization under the Sharpe ratio constraint and no-good-deal
pricing, which was popularized by \citet{cr} and \citet{bjork}. \citet{leitner} deals with an infinitesimal mean-variance risk measure of the surplus and a robust expectation of the terminal surplus under model ambiguity. He finds the hedging strategies which minimize both risk measures and the prices which make the risk measures vanish. \citet{leitner} shows that both strategies and prices coincide in a diffusion model with a non-tradeable risk factor. Finally, \citet{delong2011} considers a general combined financial and insurance model. He derives the optimal hedging strategy and the optimal price by minimizing the infinitesimal mean-variance risk measure of the surplus and by setting the infinitesimal Sharpe ratio of the surplus at a predefined level. \citet{delong2011} also shows that the optimal strategies coincide with the strategies derived under no-good-deal pricing and pricing and hedging under model ambiguity. We point out that none of the above papers covers the case of a regime-switching financial market. We are aware that \citet{donely} finds a no-good-deal price of a contingent claim in a regime-switching financial model. However, she considers a Markovian dynamics of the stock without jumps and without the feedback effect. She also does not investigate the optimal hedging strategy which can be derived by using instantaneous mean-variance hedging or hedging under model ambiguity. Consequently, to the best of our knowledge the complete characterization of the optimal price and the optimal hedging strategy under instantaneous mean-variance hedging and instantaneous Sharpe ratio pricing (no-good-deal pricing and robust pricing and hedging under model ambiguity) in a general non-Markovian regime-switching model is still missing. This paper fills this gap.

In this paper we focus on pricing and hedging of the financial risk in a general non-Markovian regime-switching financial model. More precisely, we price and hedge the financial risk induced by the Brownian motion and the multivariate counting process, which are the driving processes in our model. However, we can easily apply our results to price and hedge risks in models with non-financial risk factors. In this paper we also consider the non-Markovian regime-switching financial model and a mortality-contingent claim with a financial component (an equity-linked insurance claim). Such a model is of great importance for long-term integrated financial and insurance risk management. We use the instantaneous mean-variance risk measure and the instantaneous Sharpe ratio to price and hedge the financial and the insurance risk. To the best of our knowledge, the price and the hedging strategy which we derive in our combined financial and insurance model are new. We believe that the results of this paper are also very useful for actuarial applications.

We apply Backward Stochastic Differential Equations (BSDEs) to solve our optimization problems. The techniques are similar to the one used in \citet{delong2011}. However, some non-trivial modifications are introduced since our stock price dynamics is not continuous. We characterize the optimal price and the optimal hedging strategy with a unique solution to a nonlinear, Lipschitz BSDE with jumps. It is known that a measure solution (an arbitrage-free representation of the price) may not exist and a comparison principle (monotonicity of the pricing operator) may fail if a BSDE with jumps is used for pricing, see \cite{buckdahn}, \cite{delong2011}, \cite{royer}. However, we manage to provide simple and sufficient for financial applications conditions under which the optimal price is arbitrage-free and monotone with respect to the terminal claim and the Sharpe ratio. We also interpret the optimal hedging strategy as a delta-hedging strategy with a correction term reflecting the asymmetry of the mean-variance objective and the use of the expected profit in the objective.

This paper is structured as follows. In Section 2 we introduce the regime-switching financial model. In Section 3 we describe our pricing and hedging approach and we provide an additional motivation for the instantaneous mean-variance hedging and instantaneous Sharpe ratio pricing by giving a link to no-good-deal pricing and robust pricing and hedging under model ambiguity. In Section 4 we solve our optimization problem. Key properties of the optimal hedging strategy and the optimal price are investigated in Section 5. Pricing and hedging of insurance claims is discussed in Section 6. A numerical example is presented in Section 7.

\section{The regime-switching financial model}

\indent We deal with a probability space
$(\Omega,\mathcal{F},\mathbb{P})$ with a filtration
$\mathcal{F}=(\mathcal{F}_{t})_{0\leq t\leq T}$ and a finite time
horizon $T<\infty$. We assume that $\mathcal{F}$
satisfies the usual hypotheses of completeness ($\mathcal{F}_{0}$
contains all sets of $\mathbb{P}$-measure zero) and right continuity
($\mathcal{F}_{t}=\mathcal{F}_{t+}$). On the probability space $(\Omega,\bfi,\mathbb{P})$ we define an $\bfi$-adapted Brownian motion $W=(W(t), 0 \leq t\leq T)$ and an $\bfi$-adapted multivariate counting process $N=(N_1(t),...,N_I(t), 0\leq t\leq T)$.

We consider an economy which can be in one of $I$ states (regimes) and switches between those states randomly. For $i=1,...,I$, the counting process $N_i$ counts the number of transitions of the economy into the state $i$. We assume that
\begin{itemize}
\item[(A1)] the counting process $N_i$ has intensity $\lambda_i(t)$ where $\lambda_i:\Omega\times[0,T]\rightarrow[0,\infty)$ is an $\bfi$-predictable, bounded process.
\end{itemize}
Consequently, the compensated counting process
\beqo
\tilde{N}_i(t)=N_i(t)-\int_0^t\lambda_i(s)ds,\quad 0\leq t\leq T, \ i=1,...,I,
\eeqo
is an $\bfi$-martingale, see Chapters XI.1 and XI.4 in \cite{Y}. We remark that $\lambda_i(t)$ is an intensity of the transition of the economy into state $i$ at time $t$. Furthermore, let $J=(J(t), 0\leq t\leq T)$ denote an $\bfi$-adapted process which indicates the current state of the economy. If the economy is in a regime $k\in\{1,...,I\}$ at the initial point of time, then the dynamics of the process $J$ is given the stochastic differential equation
\beqo
dJ(t)=\sum_{i=1}^I(i-J(t-))dN_i(t),\quad J(0)=k\in\{1,...,I\},
\eeqo
A time-homogeneous Markov process $J$ arises if we choose $\lambda_i(t)=\lambda_i(J(t-))$.

The financial market consists of a risk-free bank account and a risky asset. The dynamics of the risk-free bank account $B=(B(t), 0\leq t\leq T)$ is given by the differential equation
\beq\label{bondregimes}
\frac{dB(t)}{B(t)}=r(t)dt,\quad B(0)=1,
\eeq
and the dynamics of the risky asset $S=(S(t), 0\leq t\leq T)$ is described by the stochastic differential equation
\beq\label{stockregimes}
\frac{dS(t)}{S(t-)}=\mu(t)dt+\sigma(t)dW(t)+\sj \gamma_i(t)d\tilde{N}_i(t),\quad S(0)=1.
\eeq
We assume that
\begin{itemize}
\item[(A2)] $r,\mu,\sigma,(\gamma_i)_{i=1,...,I}:\Omega\times[0,T]\rightarrow \br$ are $\bfi$-predictable, bounded processes such that there exists a unique solution $S$ to \eqref{stockregimes}. Moreover,
\beqo
&&\mu(t)\geq r(t), \quad 0\leq t\leq T,\\
&&|\delta(t)|^2=\var\geq \epsilon>0, \quad 0\leq t \leq T, \\
&&\gamma_i(t)>-1, \quad 0\leq t\leq T,\ i=1,...,I.
\eeqo
\end{itemize}
These conditions are standard in financial modelling. The first condition is clear. The second condition is a non-degeneracy condition for the volatility of the risky asset return. The third condition guarantees that the price process $S$, which solves \eqref{stockregimes}, is strictly positive, see Theorem 4.61 in \citet{jacod}. We point out that we deal with a non-Markovian model. All parameters of the model \eqref{bondregimes}-\eqref{stockregimes} are driven by the Brownian motion and the multivariate counting process. The interest rate, drift, volatility, jump amplitudes, intensities fluctuate and they depend on the past and current conditions of the economy and the financial market. Let us also notice that if the economy remains in a state, then the dynamics of the risky asset is continuous. However, if a transition into a different state occurs, then the value of the risky asset changes in a discontinuous way at the time of the transition. In the sequel we use the following notation for the instantaneous variance of the risky asset return
\beqo
|\delta(t)|^2=\var,
\eeqo
and for the instantaneous Sharpe ratio of the risky asset
\beqo
\theta(t)=\frac{\mu(t)-r(t)}{\delta(t)}.
\eeqo

The most important and practically relevant example of the financial model \eqref{bondregimes}-\eqref{stockregimes} arises when the coefficients $r,\mu,\sigma,\gamma_i$ depend only on the current state of the economy and the intensities $\lambda_i$ depend on the current state of the economy and the current value of the risky asset. In that case we investigate the dynamics
\beq\label{bsregimes}
\frac{dB(t)}{B(t)}&=&r(J(t-))dt,\nonumber\\
\frac{dS(t)}{S(t-)}&=&\mu(J(t-))dt+\sigma(J(t-))dW(t)+\sj \gamma_i(J(t-))d\tilde{N}_i(t),
\eeq
where the counting process $N_i$ has intensity $\lambda_i(t)=\lambda_i(J(t-),S(t-))$ and $J$ indicates the current state of the economy. Such a model is called a Markov-regime-switching since $(S,J)$ is a Markov process. We remark that $\lambda_i(t)=\lambda_i(J(t-),S(t-))$ denotes an intensity of the transition into state $i$ at time $t$ given the economy is in state $J(t-)$ and the stock price equals $S(t-)$. The dependence $\lambda_i(t)=\lambda_i(J(t-),S(t-))$ models the so-called feedback effect in the market, see \citet{elliot}. The complete probabilistic description of regime-switching models can be found in \cite{crepeymodel}.

\section{Instantaneous mean-variance hedging and instantaneous Sharpe ratio pricing}

Let $\xi$ be a contingent claim in the regime-switching financial market \eqref{bondregimes}-\eqref{stockregimes} which has to be covered at time $T$. We are interested in finding a hedging strategy and a price of the claim $\xi$.

Let $\pi=(\pi(t), 0\leq t\leq T)$ denote a hedging strategy, i.e. the amount of wealth which is invested into the risky asset. We introduce the set of admissible hedging strategies.
\begin{df}
A strategy $\pi:=(\pi (t), 0\leq t\leq T)$ is called admissible, written
$\pi\in\mathcal{A}$, if it satisfies the conditions:
\begin{enumerate}
    \item $\pi:[0,T]\times \Omega\rightarrow \mathbb{R}$ is an $\bfi$-predictable process,
    \item $\mathbb{E}\big[\int_{0}^{T}\big|\pi(t)\big|^{2}dt\big]<\infty$.
\end{enumerate}
\end{df}
The price of the claim is modelled as a solution to a Backward Stochastic Differential Equation (BSDE). We assume that the price process $Y:=(Y(t), 0\leq t\leq T)$ of the claim $\xi$ solves the BSDE
\beq\label{bsdeo}
Y(t)&=&\xi+\int_t^T\big(-Y(s-)r(s)-f(s)\big)ds\nonumber\\
&&-\int_t^TZ(s)dW(s)-\int_t^T\sj U_i(s)d\tilde{N}_i(s),\quad 0\leq t\leq T,
\eeq
where $f$ is the generator of the equation which has to be determined. If we decide on the form of the generator $f$, then the price of the claim $\xi$ can be uniquely defined. In order to determine the generator $f$, we use instantaneous mean-variance hedging and instantaneous Sharpe ratio pricing.

First, we define the hedging portfolio. The dynamics of the hedging portfolio $X^\pi:=(X^\pi(t),0\leq t \leq T)$ under an admissible hedging strategy $\pi\in\mathcal{A}$ is given by the stochastic differential equation
\beqo
dX^\pi(t)&=&\pi(t)\big(\mu(t)dt+\sigma(t)dW(t)+\sj \gamma_i(t)d\tilde{N}_i(t)\big)\\
&&+(X^\pi(t-)-\pi(t))r(t)dt,\\
X^\pi(0)&=&x.
\eeqo
Next, we define the surplus process $S^\pi(t)=X^\pi(t)-Y(t), \ 0\leq t\leq T$, which models the profit or the loss of the hedger resulting from the past investment and the future liability. The surplus process can also be called a hedging error. The dynamics of the surplus $S^\pi$ is described by the stochastic differential equation
\beqo
dS^\pi(t)&=&\big(\pi(t)(\mu(t)-r(t))+S^\pi(t-)r(t)-f(t)\big)dt+(\pi(t)\sigma(t)-Z(t))dW(t)\\
&&+\sj(\pi(t)\gamma_i(t)-U_i(t))d\tilde{N}_i(t).
\eeqo
By standard properties of stochastic integrals, see Theorems II.20, 28, 29, 39 in \citet{Protter}, we can derive the expected infinitesimal return on the surplus
\beq\label{generator}
\lefteqn{\be\big[dS^\pi(t)-S^\pi(t-)r(t)dt|\bfi_{t-}\big]}\nonumber\\
&&=\pi(t)(\mu(t)-r(t))dt-f(t)dt,\quad 0<t\leq T,
\eeq
and the expected infinitesimal quadratic variation of the surplus
\beq\label{quadraticvariation}
\lefteqn{\be\big[d[S^\pi,S^\pi](t)|\bfi_{t-}\big]=|\pi(t)\sigma(t)-Z(t)|^2dt}\nonumber\\
&&+\sj|\pi(t)\gamma_i(t)-U_i(t)|^2\lambda_i(t)dt,\quad 0<t\leq T.
\eeq
Our goal is to find an admissible hedging strategy $\pi\in\mathcal{A}$ which minimizes the instantaneous mean-variance risk of the surplus
\beq\label{objectivestrategy}
\rho(S^\pi)=L(t)\sqrt{\be\big[d[S^\pi,S^\pi](t)|\bfi_{t-}\big]/dt}-\big(\be\big[dS^\pi(t)-S^\pi(t-)r(t)dt|\bfi_{t-}\big]/dt\big),
\eeq
for all $t\in(0,T]$, and set the price of $\xi$ (find the generator $f$ of the BSDE \eqref{bsdeo}) in such a way that the instantaneous Sharpe ratio of the surplus equals a predefined target $L$, i.e.
\beq\label{objectiveprice}
\frac{\be\big[dS^\pi(t)-S^\pi(t-)r(t)dt|\bfi_{t-}\big]/dt}{\sqrt{\be\big[d[S^\pi,S^\pi](t)|\bfi_{t-}\big]/dt}}=L(t),
\eeq
for all $t\in(0,T]$. The hedging and pricing objectives \eqref{objectivestrategy}-\eqref{objectiveprice} are called the instantaneous mean-variance hedging and instantaneous Sharpe ratio pricing. We shall assume that
\begin{itemize}
\item[(A3)] $L$ is an $\bfi$-predictable process such that $L(t)\geq \theta(t)+\epsilon, \ 0\leq t\leq T, \ \epsilon>0$.
\end{itemize}
Since the Sharpe ratio of the surplus $L$ is an $\bfi$-predictable process, it can depend on the economy and the financial market. In particular, the hedger may use different Sharpe ratios in different states of the economy. Such an assumption is important from the practical point of view since investors have different profit expectations in a bull market and in a bear market. We also require that the Sharpe ratio of the surplus is strictly greater than the Sharpe ratio of the risky asset. Such an assumption is obvious since the hedger trading $\xi$ would require a Sharpe ratio $L$ which is strictly greater than the Sharpe ratio $\theta$ which can be earned by simply investing in the stock $S$.

\citet{youngfin}, \citet{younglife} and \citet{youngannuity} have advocated the instantaneous mean-variance hedging and instantaneous Sharpe ratio pricing for hedging and pricing financial and insurance risks. Let us remark that the hedging and pricing objectives \eqref{objectivestrategy}-\eqref{objectiveprice} are easy to communicate, are based on the first two moments of the hedging error, are related to the Markowitz portfolio selection problem and involve a Sharpe ratio which is well understood by investors. These four features already make the instantaneous mean-variance hedging and instantaneous Sharpe ratio pricing a reasonable method for pricing and hedging risks in incomplete markets. Interestingly, the instantaneous mean-variance hedging and instantaneous Sharpe ratio pricing can be related to no-good-deal pricing and robust pricing and hedging under model ambiguity.

It turns out that the price derived under the instantaneous Sharpe ratio pricing \eqref{objectiveprice} is equivalent to the price derived under no-good-deal pricing, see \citet{youngannuity}, \citet{youngfin}, \citet{delong2011}, \citet{younglife}. Hence, the theory of no-good-deal pricing gives us an additional motivation and justification for the instantaneous mean-variance hedging and instantaneous Sharpe ratio pricing. The no-good-deal price of the claim $\xi$ is defined as a solution to the following optimization problem
\beq\label{nogooddealobjective}
Y(t)=\sup_{(\psi,\phi)\in\mathcal{Q}}\be^{\bq^{\psi,\phi}}\big[e^{-\int_t^Tr(s)ds}\xi|\bfi_t\big],\quad 0\leq t\leq T,
\eeq
where $\bq^{\psi,\phi}$ is an equivalent martingale measure. Under no-good-deal pricing we price a claim with a least favorable pricing measure from a set of equivalent martingale measures. The set of equivalent martingale measures is defined by the Radon-Nikodym derivative
\beqo
\frac{d\bq^{\psi,\phi}}{d\bp}\Big|\bfi_t=M^{\psi,\phi}(t),\quad 0\leq t\leq T,\quad (\psi,\phi)\in\mathcal{Q},
\eeqo
where
\beqo
\frac{dM^{\psi,\phi}(t)}{M^{\psi,\phi}(t-)}=-\psi(t)dW(t)-\sj\phi_i(t)d\tilde{N}_i(t),\quad M(0)=1,\quad (\psi,\phi)\in\mathcal{Q},
\eeqo
and
\beqo
\mathcal{Q}&=&\Big\{\bfi-predictable \ processes \ (\psi,\phi)=(\psi,\phi_1,...,\phi_I) \ such \ that\nonumber\\
&&|\psi(t)|^2+\sj|\phi_i(t)|^2\lambda_i(t)\leq |L(t)|^2,\nonumber\\
&&\psi(t)\sigma(t)+\sj \phi_i(t)\gamma_i(t)\lambda_i(t)=\mu(t)-r(t),\nonumber\\
&&\phi_i(t)<1, \quad 0\leq t\leq T,\ i=1,...,I\Big\}.
\eeqo
Let us briefly explain the conditions from the set $\mathcal{Q}$. The third condition is clear as it guarantees that $M^{\psi,\phi}$ is strictly positive, see Theorem 4.61 in \citet{jacod}. Recalling the Girsanov's theorem, see Theorem III.40 in \cite{Protter}, we can derive the dynamics of the stock
\beqo
\frac{dS(t)}{S(t-)}&=&\big(\mu(t)-\psi(t)\sigma(t)-\sj \phi_i(t)\gamma_i(t)\lambda_i(t)\big)dt\\
&&+\sigma(t)dW^{\bq^{\psi,\phi}}(t)+\sj\gamma_i(t)d\tilde{N}^{\bq^{\psi,\phi}}(t),
\eeqo
and we can observe that the second condition implies that the discounted stock process is a $\bq^{\psi,\phi}$-martingale for any $(\phi,\psi)\in\mathcal{Q}$ and $M^{\phi,\psi}$ defines a set of equivalent martingale measures for the market \eqref{bondregimes}-\eqref{stockregimes}. Finally, by the Girsanov's theorem and standard arguments for BSDEs, see Proposition 2.2 in \citet{karouipengbsde}, we deduce that any arbitrage-free price process $Y^{\psi,\phi}(t)=\be^{\bq^{\psi,\phi}}\big[e^{-\int_t^Tr(s)ds}\xi|\bfi_t\big], \ 0\leq t\leq T, \ (\psi,\phi)\in\mathcal{Q},$ has the dynamics
\beq\label{bsdea}
dY^{\psi,\phi}(t)&=&Y^{\psi,\phi}(t-)r(t)dt+Z^{\psi,\phi}(t)\psi(t)dt+\sj U^{\psi,\phi}_i(t)\phi_i(t)\lambda_i(t)dt\nonumber\\
&&+Z^{\phi,\psi}(t)dt+\sj U^{\psi,\phi}_i(t)d\tilde{N}_i(t),\nonumber\\
Y^{\psi,\phi}(T)&=&\xi.
\eeq
We can derive the bound for the instantaneous Sharpe ratio of the arbitrage-free price process $Y^{\psi,\phi}$ of $\xi$:
\beqo
\Big|\frac{\be\big[dY^{\psi,\phi}(t)-Y^{\psi,\phi}(t-)r(t)dt|\bfi_{t-}\big]/dt}{\sqrt{\be\big[d[Y^{\psi,\phi},Y^{\psi,\phi}](t)|\bfi_{t-}\big]/dt}}\Big|&=&\frac{\big|Z^{\psi,\phi}(t)\psi(t)+\sj U^{\psi,\phi}_i(t)\phi_i(t)\lambda_i(t)\big|}{\sqrt{|Z^{\psi,\phi}(t)|^2+\sj |U^{\psi,\phi}_i(t)|^2\lambda_i(t)}}\\
&\leq&\sqrt{|\psi(t)|^2+\sj|\phi_i(t)|^2\lambda_i(t)},
\eeqo
and we conclude that the first condition in $\mathcal{Q}$ implies that the instantaneous Sharpe ratio of an arbitrage-free price process of the claim $\xi$ is bounded by $L$. The process $L$ defines a so-called no-good-deal range in the financial market and it represents the bound on possible gains in the financial market measured by the instantaneous Sharpe ratio. The existence of such a maximal gain is justified by empirical financial data, see \cite{cr}, \cite{bjork} for motivation. Hence, under the no-good-deal pricing \eqref{nogooddealobjective} we price the claim $\xi$ with a least favorable pricing measure under the Sharpe ratio constraint which excludes too high (and unrealistic) gains which could be earned (but only theoretically) by writting the contract with an arbitrary high price. We remark that by the least favorable pricing measure we mean a measure which leads to the highest expected pay-off from the claim.

The price and the hedging strategy derived under the instantaneous mean-variance hedging and instantaneous Sharpe ratio \eqref{objectivestrategy}-\eqref{objectiveprice} also coincide with the price and the hedging strategy derived under robust pricing and hedging under model ambiguity, see \cite{leitner}, \cite{delong2011}, \cite{pelsser}. The idea behind robust pricing and hedging under model ambiguity gives us another motivation and justification for the instantaneous mean-variance hedging and instantaneous Sharpe ratio pricing. Let us introduce a set which consists of equivalent measures defined by the Radon-Nikodym derivative
\beqo
\frac{d\bq^{\psi,\phi}}{d\bp}\Big|\bfi_t=M^{\psi,\phi}(t),\quad 0\leq t\leq T,\quad (\psi,\phi)\in\mathcal{P},
\eeqo
and
\beqo
\frac{dM^{\psi,\phi}(t)}{M^{\psi,\phi}(t-)}=-\psi(t)dW(t)-\sj\phi_i(t)d\tilde{N}_i(t),\quad M(0)=1,\quad (\psi,\phi)\in\mathcal{P},
\eeqo
\beqo
\mathcal{P}&=&\Big\{\bfi-predictable \ processes \ (\psi,\phi)=(\psi,\phi_1,...,\phi_I) \ such \ that\nonumber\\
&&|\psi(t)|^2+\sj|\phi_i(t)|^2\lambda_i(t)\leq |L(t)|^2,\nonumber\\
&&\phi_i(t)<1, \quad 0\leq t\leq T, \ i=1,...,I\Big\}.
\eeqo
We now define the price and the hedging strategy as a solution to the following robust optimization problem
\beq\label{robustproblem}
Y(t)&=&\inf_{\pi\in\mathcal{A}}\sup_{(\psi,\phi)\in\mathcal{P}}\be^{\bq^{\psi,\phi}}\Big[-\big(e^{-\int_t^Tr(s)ds}X^\pi(T)-X(t)\nonumber\\
&&\quad \ \quad \ -e^{-\int_t^Tr(s)ds}F\big)\big|\bfi_t\Big],\quad 0\leq t\leq T.
\eeq
The set $\mathcal{P}$ represents different beliefs (different assumptions) about the parameters or the evolution of the risk factors in our model. One way of determining the set $\mathcal{P}$ for ambiguity modelling is to specify confidence sets around the estimates of the parameters and to take for $\mathcal{P}$ the class of all measures that are consistent with these confidence sets. Then, the process $L$ can be interpreted as an estimation error. Alternatively, the elements of $\mathcal{P}$ can be interpreted as prior models which describe probabilities of future scenarios for the risk factors. Then, the process $L$ can define the range of equivalent probabilities for every scenario. Hence, under the objective of pricing and hedging under model ambiguity \eqref{robustproblem} we aim to find a hedging strategy for the claim $\xi$ which minimizes the expected shortfall in the terminal surplus under a least favorable measure describing future scenarios and we price the claim $\xi$ with a value which offsets this worst expected shortfall.

We solve the no-good-deal pricing problem \eqref{nogooddealobjective} in Section 5.1, and we see the equivalence between \eqref{nogooddealobjective} and the instantaneous Sharpe ratio pricing \eqref{objectiveprice}. The equivalence between the instantaneous mean-variance hedging and instantaneous Sharpe ratio pricing \eqref{objectivestrategy}-\eqref{objectiveprice} and the pricing and hedging under model ambiguity \eqref{robustproblem} is not proved in this paper. Details can be obtained from the authors upon the request. We remark that the proof of the equivalence can be established by modifying the steps of the proofs from \citet{Bech2} and \citet{delong2011}.

\section{The optimal price and the optimal hedging strategy}

We characterize the optimal hedging strategy and the optimal price process which solve \eqref{objectivestrategy}-\eqref{objectiveprice} with a solution to a backward stochastic differential equation. In order to use the theory of BSDEs, we assume that the weak property of predictable representation holds, see Proposition 7.5 in \cite{crepeymodel} and Chapter XIII.2 in \cite{Y}, i.e.
\begin{itemize}
\item [(A4)] every $(\mathbb{P},\mathcal{F})$ local martingale $M$ has the
representation
\begin{eqnarray*}
M(t)=M(0)+\int_{0}^{t}\mathcal{Z}(s)dW(s)+\int_0^t\sj\mathcal{U}_i(s)d\tilde{N}_i(s)\quad 0\leq
t\leq T,
\end{eqnarray*}
with $\bfi$-predictable processes $(\mathcal{Z}, \mathcal{U}_1, ...,\mathcal{U}_I)$ which are integrable in the It\^{o} sense.
\end{itemize}
This assumption is satisfied if we define the probability space and the driving processes in a appropriate way, see \citet{Bech} and \cite{crepeymodel}.

We present the main theorem of this paper.
\begin{thm}\label{pricehedge}
We investigate the instantaneous mean-variance hedging and instantaneous Sharpe ratio pricing \eqref{objectivestrategy}-\eqref{objectiveprice} of the claim $\xi$. Let $\xi$ be an $\bfi$-measurable claim such that $\be[|\xi|^2]<\infty$, and assume that (A1)-(A4) hold. Consider the BSDE
\beq\label{bsde1}
\lefteqn{Y(t)=\xi+\int_t^T\Big(-Y(s)r(s)-\frac{Z(s)\sigma(s)+\sj U_i(s)\gamma_i(s)\lambda_i(s)}{\delta(s)}\theta(s)}\nonumber\\
&&+\sqrt{|L(s)|^2-|\theta(s)|^2}\nonumber\\
&&\quad \cdot\sqrt{|Z(s)|^2+\sj |U_i(s)|^2\lambda_i(s)-\frac{|Z(s)\sigma(s)+\sj U_i(s)\gamma_i(s)\lambda_i(s)|^2}{|\delta(s)|^2}} \ \Big)ds\nonumber\\
&&-\int_t^TZ(s)dW(s)-\int_t^T\sj U_i(s)d\tilde{N}_i(s),\quad 0\leq t\leq T,
\eeq
with its unique solution $(Y,Z,U_1,...,U_I)$. The optimal admissible hedging strategy $\pi^*\in\mathcal{A}$ for $\xi$ is of the form
\beq\label{strategy1}
\lefteqn{\pi^*(t)=\frac{Z(t)\sigma(t)+\sj U_i(t)\gamma_i(t)\lambda_i(t)}{|\delta(t)|^2}}\nonumber\\
&&+\frac{\theta(t)}{\delta(t)\sqrt{|L(t)|^2-|\theta(t)|^2}}\nonumber\\
&&\quad\cdot\sqrt{|Z(t)|^2+\sj |U_i(t)|^2\lambda_i(t)-\frac{|Z(t)\sigma(t)+\sj U_i(t)\gamma_i(t)\lambda_i(t)|^2}{|\delta(t)|^2}},\quad 0\leq t\leq T,\quad \ \quad \
\eeq
and the price process of $\xi$ is given by $Y$.
\end{thm}
\Proof
\emph{Step 1)} First, we find the optimal solution to our optimization problem \eqref{objectivestrategy}. By \eqref{generator} and \eqref{quadraticvariation} we have to find a minimizer of the function
\beqo
h(\pi)=L\sqrt{|\pi\sigma-z|^2+\sj|\pi\gamma_i-u_i|^2\lambda_i}-\pi(\mu-r).
\eeqo
Since (A3) holds, then $\lim_{\pi\rightarrow+\infty}h(\pi)=+\infty$ and $\lim_{\pi\rightarrow-\infty}h(\pi)=+\infty$. Consequently, there exists an odd number of extreme points of $h$ and at least one minimizer of $h$. We can notice that the function $h$ is differentiable everywhere, except at $\pi=\frac{z}{\sigma}$ if $\frac{z}{\sigma}=\frac{u_i}{\gamma_i}$. Hence, let us find stationary points of $h$ by solving the equation
\beq\label{derivativeoptimum}
0=h'(\pi)=L\frac{(\pi\sigma-z)\sigma+\sj(\pi\gamma_i-u_i)\gamma_i\lambda_i}{\sqrt{|\pi\sigma-z|^2+\sj|\pi\gamma_i-u_i|^2\lambda_i}}-(\mu-r).
\eeq
Given that the stationary point $\pi$ must satisfy
\beq\label{quadraticcondition}
0\leq (\pi\sigma-z)\sigma+\sj(\pi\gamma_i-u_i)\gamma_i\lambda_i=\pi(\sigma^2+\sj\gamma_i^2\lambda_i)-z\sigma-\sj u_i\gamma_i\lambda_i,
\eeq
we end up with the quadratic equation
\beqo
\frac{(\mu-r)^2}{L^2}\big(|\pi\sigma-z|^2+\sj|\pi\gamma_i-u_i|^2\lambda_i\big)=\big|(\pi\sigma-z)\sigma+\sj(\pi\gamma_i-u_i)\gamma_i\lambda_i\big|^2,
\eeqo
which after easy, but tedious, calculations reduces to
\beq\label{quadraticeq}
\lefteqn{\pi^2\big(\sigma^2+\sj\gamma_i\lambda_i\big)\big(\big(\frac{\mu-r}{L}\big)^2-\sigma^2-\sj\gamma_i\lambda_i\big)}\nonumber\\
&&+2\pi\big(z\sigma+\sj u_i\gamma_i\lambda_i\big)\big(\sigma^2+\sj\gamma_i\lambda_i-\big(\frac{\mu-r}{L}\big)^2\big)\nonumber\\
&&+\big(\frac{\mu-r}{L}\big)^2\big(z^2+\sj u_i^2\lambda_i\big)-\big(z\sigma+\sj u_i\gamma_i\lambda_i\big)^2=0.
\eeq
We can calculate
\beqo
\triangle&=&4\Big(\sigma^2+\sj\gamma_i^2\lambda_i-\big(\frac{\mu-r}{L}\big)^2\Big)\big(\frac{\mu-r}{L}\big)^2\big(\sigma^2+\sj\gamma_i^2\lambda_i\big)\\
&&\quad\cdot\Big(z^2+\sj u_i^2\lambda_i-\frac{(z\sigma+\sj u_i\gamma_i\lambda_i)^2}{\sigma^2+\sj\gamma_i^2\lambda_i}\Big),
\eeqo
and we obtain that the quadratic equation \eqref{quadraticeq} has two roots
\beqo
\lefteqn{\pi_1^*=\frac{z\sigma+\sj u_i\gamma_i\lambda_i}{\sigma^2+\sj\gamma_i^2\lambda_i}}\\
&&+\frac{\mu-r}{(\sigma^2+\sj\gamma_i^2\lambda_i)\sqrt{L^2-\frac{(\mu-r)^2}{\sigma^2+\sj\gamma_i^2\lambda_i}}}\sqrt{z^2+\sj u_i^2\lambda_i-\frac{(z\sigma+\sj u_i\gamma_i\lambda_i)^2}{\sigma^2+\sj\gamma_i^2\lambda_i}},\\
\lefteqn{\pi_2^*=\frac{z\sigma+\sj u_i\gamma_i\lambda_i}{\sigma^2+\sj\gamma_i^2\lambda_i}}\\
&&-\frac{\mu-r}{(\sigma^2+\sj\gamma_i^2\lambda_i)\sqrt{L^2-\frac{(\mu-r)^2}{\sigma^2+\sj\gamma_i^2\lambda_i}}}\sqrt{z^2+\sj u_i^2\lambda_i-\frac{(z\sigma+\sj u_i\gamma_i\lambda_i)^2}{\sigma^2+\sj\gamma_i^2\lambda_i}}.
\eeqo
It is straightforward to check that only $\pi_1^*$ satisfies \eqref{quadraticcondition}. By the properties of $h$ we can now conclude that $\pi^*_1$ is the unique minimizer of the function $h$. From \eqref{objectiveprice} we immediately deduce that the optimal generator $f^*$ of the BSDE is given by the formula
\beqo
f^*=\pi^*_1(\mu-r)-L\sqrt{|\pi^*_1\sigma-z|^2+\sj|\pi^*_1\gamma_i-u_i|^2\lambda_i},
\eeqo
and recalling \eqref{derivativeoptimum} we derive
\beqo
f^*=\pi^*_1(\mu-r)-\frac{L^2}{\mu-r}\Big((\pi^*_1\sigma-z)\sigma+\sj(\pi^*_1\gamma_i-u_i)\gamma_i\lambda_i\Big).
\eeqo
Substituting $\pi_1^*$, we obtain the generator of our BSDE. \\
\emph{Step 2)} We prove the existence of a unique solution to the BSDE \eqref{bsde1}. We can notice that the strategy
\beq\label{varianceoptimum}
\tilde{\pi}^*(t,Z(t),U(t))=\frac{Z(t)\sigma(t)+\sj U_i(t)\gamma_i(t)\lambda_i(t)}{|\delta(t)|^2},
\eeq
is the unique minimizer of the quadratic variation of the surplus \eqref{quadraticvariation} and we have
\beq\label{representationobjective}
\lefteqn{\sqrt{|\tilde{\pi}^*(t,Z(t),U(t))\sigma(t)-Z(t)|^2+\sj|\tilde{\pi}^*(t,Z(t),U(t))\gamma_i(t)-U_i(t)|^2\lambda_i(t)}}\nonumber\\
&=&\sqrt{|Z(t)|^2+\sj |U_i(t)|^2\lambda_i(t)-\frac{|Z(t)\sigma(t)+\sj U_i(t)\gamma_i(t)\lambda_i(t)|^2}{|\delta(t)|^2}}.
\eeq
We can now show that the generator $f$ of the BSDE \eqref{bsde1} is Lipschitz continuous in the sense that
\beqo
\lefteqn{|f(t,Y(t),Z(t),U(t))-f(t,Y'(t),Z'(t),U'(t))|^2}\\
&=&\Big|Y(t)r(t)-Y'(t)r(t)\\
&&+\tilde{\pi}^*(t,Z(t),U(t))(\mu(t)-r(t))-\tilde{\pi}^*(t,Z'(t),U'(t))(\mu(t)-r(t))\\
&&-\sqrt{|L(t)|^2-|\theta(t)|^2}\\
&&\quad \cdot\sqrt{|\tilde{\pi}^*(t,Z(t),U(t))\sigma(t)-Z(t)|^2+\sj|\tilde{\pi}^*(t,Z(t),U(t))\gamma_i(t)-U_i(t)|^2\lambda_i(t)}\\
&&+\sqrt{|L(t)|^2-|\theta(t)|^2}\\
&&\quad\cdot\sqrt{|\pi^*(t,Z'(t),U'(t))\sigma(t)-Z'(t)|^2+\sj|\pi^*(t,Z'(t),U'(t))\gamma_i(t)-U'_i(t)|^2\lambda_i(t)} \Big|^2\\
&\leq& K\Big(|\tilde{\pi}^*(t,Z(t),U(t))-\tilde{\pi}^*(t,Z'(t),U'(t))|^2\\
&&+|Y(t)-Y'(t)|^2+|Z(t)-Z'(t)|^2+\sj|U_i(t)-U'_i(t)|^2\lambda_i(t)\Big)\\
&\leq& K\Big(|Y(t)-Y'(t)|^2+|Z(t)-Z'(t)|^2+\sj|U_i(t)-U'_i(t)|^2\lambda_i(t)\Big),
\eeqo
where we use the representation \eqref{representationobjective}, the boundedness assumptions (A2) and the inequality
\beq\label{norminequality}
|\sqrt{x^2+a^2}-\sqrt{y^2+b^2}|^2\leq |x-y|^2+|a-b|^2.
\eeq
By standard results on BSDEs there exists a unique solution to the BSDE \eqref{bsde1}, see Proposition 3.2 in \cite{Bech}.\\
\emph{Step 3)} We are left with showing the admissability of the optimal strategy. Standard results on BSDEs, see Proposition 3.2 in \cite{Bech}, yield that $Y$ is $\bfi$-adapted, $(Z,U_1,...,U_I)$ are $\bfi$-predictable, and
\beqo
&&\be\big[\sup_{t\in[0,T]}|Y(t)|^2\big]<\infty, \quad \be\Big[\int_0^T|Z(s)|^2ds\Big]<\infty,\\
&&\be\Big[\sj\int_0^T|U_i(s)|^2\lambda_i(s)ds\Big]<\infty.
\eeqo
Hence, it is straightforward to conclude that $\pi^*\in\mathcal{A}$.
\cbdu

We have succeeded in characterizing the optimal hedging strategy and the optimal price process with a unique solution to a nonlinear BSDE which has a Lipschitz generator.

\section{Properties of the price and the hedging strategy}

In this section we investigate some important properties of the optimal price process and the optimal hedging strategy.

\subsection{The arbitrage-free representation of the price and no-good-deal pricing}

From the point of view of the arbitrage-free pricing theory, the crucial point is to check whether the price process \eqref{bsde1} can be represented in the form
\beqo
Y(t)=\be^\bq\Big[e^{-\int_t^Tr(s)ds}\xi|\bfi_t\Big],\quad 0\leq t\leq T,
\eeqo
where $\bq$ is an equivalent martingale measure. Looking at the BSDE \eqref{bsde1}, it is rather difficult to guess the form of the equivalent martingale measure and prove the arbitrage-free representation. Since no-good-deal pricing aims at finding a worst equivalent martingale measure for arbitrage-free pricing, we now solve the no-good-deal pricing problem \eqref{nogooddealobjective}. As a by-product, we obtain the arbitrage-free representation of the price \eqref{bsde1}.

\begin{thm}\label{nogooddealprice}
Let $\xi$ be an $\bfi$-measurable claim such that $\be[|\xi|^2]<\infty$, and assume that (A1)-(A4) hold. Consider the BSDE
\beq\label{bsdearbitrage}
\lefteqn{Y(t)=\xi+\int_t^T\Big(-Y(s)r(s)-\frac{Z(s)\sigma(s)+\sj U_i(s)\gamma_i(s)\lambda_i(s)}{\delta(s)}\theta(s)}\nonumber\\
&&+\sqrt{|L(s)|^2-|\theta(s)|^2}\nonumber\\
&&\quad \cdot\sqrt{|Z(s)|^2+\sj |U_i(s)|^2\lambda_i(s)-\frac{|Z(s)\sigma(s)+\sj U_i(s)\gamma_i(s)\lambda_i(s)|^2}{|\delta(s)|^2}} \ \Big)ds\nonumber\\
&&-\int_t^TZ(s)dW(s)-\int_t^T\sj U_i(s)d\tilde{N}_i(s)\quad 0\leq t\leq T,
\eeq
with its unique solution $(Y,Z,U_1,...,U_I)$. Let
\beq\label{arbitragecondition1}
\lambda_i(t)>0\Longrightarrow U_i(t)\neq 0, \quad 0\leq t\leq T, \ i=1,...,I,
\eeq
and
\beq\label{arbitragecondition2}
\sqrt{|L(t)|^2-|\theta(t)|^2}+\frac{|\gamma_i(t)|\sqrt{\lambda_i(t)}}{\delta(t)}\theta(t)<\sqrt{\lambda_i(t)},\quad 0\leq t\leq T, \ i=1,...,I,
\eeq
on the set $\{\lambda_i(t)>0\}$. The optimal equivalent martingale measure $\bq^{\psi^*,\phi^*}$ which solves the optimization problem \eqref{nogooddealobjective} is determined by the processes
\beq\label{optimalmeasure}
\psi^*(t)&=&\theta(t)\mathbf{1}\{\forall_{i=1,...,I}\lambda_i(t)=0\}\nonumber\\
&&+\frac{Z(t)-\sigma(t)K^*_1(t,Z(t),U(t))}{2K^*_2(t,Z(t),U(t))}\mathbf{1}\{\exists_{i=1,...,I}\lambda_i(t)>0\},\quad 0\leq t\leq T,\nonumber\\
\phi_i^*(t)&=&\frac{U_i(t)-\gamma_i(t)K^*_1(t,Z(t),U(t))}{2K^*_2(t,Z(t),U(t))}\mathbf{1}\{\lambda_i(t)>0\},\quad 0\leq t\leq T, \ i=1,...,I.\quad \
\eeq
where
\beqo
K^*_2(t,Z(t),U(t))&=&-\frac{1}{2}\frac{\sqrt{|Z(t)|^2+\sj |U_i(t)|^2\lambda_i(t)-\frac{|Z(t)\sigma(t)+\sj U_i(t)\gamma_i(t)\lambda_i(t)|^2}{|\delta(t)|^2}}}{\sqrt{|L(t)|^2-|\theta(t)|^2}},\\
K^*_1(t,Z(t),U(t))&=&-\frac{\theta(t)}{\delta(t)}2K^*_2(t,Z(t),U(t))+\frac{Z(t)\sigma(t)+\sj U_i(t)\gamma_i(t)\lambda_i(t)}{|\delta(t)|^2}.
\eeqo
Moreover, the process $Y$ coincides with the optimal value function of the optimization problem \eqref{nogooddealobjective} and we have
\beqo
Y(t)=\sup_{(\psi,\phi)\in\mathcal{Q}}\be^{\bq^{\psi,\phi}}\big[e^{-\int_t^Tr(s)ds}\xi|\bfi_t\big]=\be^{\bq^{\psi^*,\phi^*}}\big[e^{-\int_t^Tr(s)ds}\xi|\bfi_t\big]\quad 0\leq t\leq T.
\eeqo
\end{thm}
\Proof
\emph{Step 1)} First, we solve the optimization problem
\beq\label{lagrange}
&&z\psi+\sj u_i\phi_i\lambda_i \rightarrow_{\psi, \phi_1,...,\phi_I} \min\nonumber\\
&&\psi\sigma+\sj \phi_i\gamma_i\lambda_i=\mu-r,\nonumber\\
&&|\psi|^2+\sj|\phi_i|^2\lambda_i\leq L^2.
\eeq
If $\lambda_i=0, \ i=1,...,I$, then we immediately get the optimal solution. Let us now assume w.l.o.g. that $\lambda_i>0,\ i=1,...,I$. By \eqref{arbitragecondition1} we can also assume that $u_i\lambda_i>0, \ i=1,...,I$. We can notice that $\psi=\frac{\mu-r}{\sigma^2+\sj\gamma_i^2\lambda_i}\sigma, \phi_i=\frac{\mu-r}{\sigma^2+\sj\gamma_i^2\lambda_i}\gamma_i, i=1,...,I,$ are the non-regular point of the constraints. The value of the objective function at the non-regular point is equal to
\beq\label{valuenonregular}
\frac{z\sigma+\sj u_i\gamma_i\lambda_i}{\sigma^2+\sj\gamma_i^2\lambda_i}(\mu-r).
\eeq
Let us now deal with regular points of the constraints. We introduce the Lagrangian
\beqo
F(\psi,\phi, K_1, K_2)&=&z\psi+\sj u_i\phi_i\lambda_i-K_1(\psi\sigma+\sj \phi_i\gamma_i\lambda_i-\mu-r)\\
&&-K_2(|\psi|^2+\sj|\phi_i|^2\lambda_i-L^2).
\eeqo
The first order conditions yield the set of equations
\beq\label{firstorederc}
z-K_1\sigma-2K_2\psi&=&0,\nonumber\\
u_i\lambda_i-K_1\gamma_i\lambda_i-2K_2\phi_i\lambda_i&=&0,\quad i=1,...,I,\nonumber\\
\psi\sigma+\sj \phi_i\gamma_i\lambda_i&=&\mu-r,\nonumber\nonumber\\
|\psi|^2+\sj|\phi_i|^2\lambda_i&=&L^2,
\eeq
and the second order condition gives us that the minimum in \eqref{lagrange} is attained for $K_2<0$. From \eqref{firstorederc} we easily obtain
\beq\label{firstorderc2}
\psi&=&\frac{z-K_1\sigma}{2K_2},\nonumber\\
\phi_i&=&\frac{u_i-K_1\gamma_i}{2K_2},\quad i=1,...,I,\nonumber\\
K_1&=&-\frac{\mu-r}{\sigma^2+\sj|\gamma_i|^2\lambda_i}2K_2+\frac{z\sigma+\sj u_i\gamma_i\lambda_i}{\sigma^2+\sj|\gamma_i|^2\lambda_i},\nonumber\\
4K^2_2&=&\Big|\frac{z-K_1\sigma}{L}\Big|^2+\sj\Big|\frac{u_i-K_1\gamma_i}{L}\Big|^2\lambda_i.
\eeq
Substituting the formula for $K_1$ into the last equation in \eqref{firstorderc2}, we can derive the quadratic equation
\beqo
\lefteqn{4K_2^2\Big(L^2-\frac{(\mu-r)^2}{\sigma^2+\sj|\gamma_i|^2\lambda_i}\Big)}\\
&=&\Big|z-\sigma\frac{z\sigma+\sj u_i\gamma_i\lambda_i}{\sigma^2+\sj|\gamma_i|^2\lambda_i}\Big|^2+\sj\Big|u_i-\gamma_i\frac{z\sigma+\sj u_i\gamma_i\lambda_i}{\sigma^2+\sj|\gamma_i|^2\lambda_i}\Big|^2\lambda_i.
\eeqo
Recalling \eqref{varianceoptimum}-\eqref{representationobjective} we get the optimal $K^*_2<0$. We calculate the value of the objective function \eqref{lagrange} at the regular point $(\psi^*,\phi^*)$. Using the formulas from \eqref{firstorderc2}, we derive
\beq\label{valueregular}
\lefteqn{z\psi^*+\sj \phi^*_iu_i\lambda_i=\frac{z\sigma+\sj u_i\gamma_i\lambda_i}{\sigma^2+\sj|\gamma_i|^2\lambda_i}(\mu-r)}\nonumber\\
&&-\sqrt{L^2-\Big(\frac{\mu-r}{\sigma^2+\sj|\gamma_i|^2\lambda_i}\Big)^2}\sqrt{z^2+\sj|u_i|^2\lambda_i-\frac{(z\sigma+\sj u_i\gamma_i\lambda_i)^2}{\sigma^2+\sj|\gamma_i|^2\lambda_i}}.\quad \
\eeq
Since \eqref{valueregular} is less than \eqref{valuenonregular}, we conclude that $(\psi^*,\phi^*)$ is the optimal solution to \eqref{lagrange}.
\emph{Step 2)} We now find the optimal solution to our no-good-deal optimization problem \eqref{nogooddealobjective}. We choose $(\psi,\phi)\in\mathcal{Q}$. Let $Y^{\psi,\phi}(t)=\be^{\bq^{\psi,\phi}}\big[e^{-\int_t^Tr(s)ds}\xi|\bfi_t\big], \ 0\leq t\leq T$. By standard arguments on BSDEs, see Proposition 2.2 in \cite{karouipengbsde} and Proposition 3.2 in \cite{Bech}, the process $Y^{\psi,\phi}$ can be characterized as a unique solution to the BSDE
\beq\label{bsdeaa}
dY^{\psi,\phi}(t)&=&Y^{\psi,\phi}(t)r(t)dt+Z^{\psi,\phi}(t)\psi(t)dt+\sj U^{\psi,\phi}_i(t)\phi_i(t)\lambda_i(t)dt\nonumber\\
&&+Z^{\phi,\psi}(t)dW(t)+\sj U^{\psi,\phi}_i(t)d\tilde{N}_i(t),\nonumber\\
Y^{\psi,\phi}(T)&=&\xi.
\eeq
Let us consider the process $Y$ which solves the BSDE \eqref{bsdearbitrage}. The existence of a unique solution $Y$ to \eqref{bsdearbitrage} is established in Theorem \ref{pricehedge}. By \eqref{valuenonregular} we can deal with the dynamics
\beq\label{bsdeb}
dY(t)&=&Y(t)r(t)dt+Z(t)\psi^*(t)dt+\sj U_i(t)\phi^*_i(t)\lambda_i(t)dt\nonumber\\
&&+Z(t)dW(t)+\sj U_i(t)d\tilde{N}_i(t),\nonumber\\
Y(T)&=&\xi,
\eeq
where $(\psi^*,\phi^*)$ are defined in \eqref{optimalmeasure}. By the Girsanov' theorem, see Theorem III.40 in \cite{Protter}, we derive
\beqo
\lefteqn{d(Y^{\psi,\phi}(t)-Y(t))=(Y^{\psi,\phi}(t)-Y(t))r(t)dt}\\
&&+\big(Z(t)\psi(t)+\sj U_i(t)\phi_i(t)\lambda_i(t)-Z(t)\psi^*(t)+\sj U_i(t)\phi^*_i(t)\lambda_i(t)\big)dt\\
&&+(Z^{\psi,\phi}(t)-Z(t))dW^{\bq^{\psi,\phi}}+\sj (U^{\psi,\phi}_i(t)-U_i(t))d\tilde{N}_i^{\bq^{\psi,\phi}}(t),\\
\lefteqn{Y^{\psi,\phi}(T)-Y(T)=0,}
\eeqo
where $W^{\bq^{\psi,\phi}}$ and $\tilde{N}^{\bq^{\psi,\phi}}$ are the $\bq^{\psi,\phi}$-Brownian motion and the $\bq^{\psi,\phi}$-compensated counting process. The result established in Step 1) and the classical steps used for proving the comparison for BSDEs, see Theorem 2.5 in \cite{royer}, yield that $Y^{\phi,\psi}(t)\leq Y(t), \ 0\leq t\leq T$. By \eqref{varianceoptimum},\eqref{representationobjective} and \eqref{firstorderc2} we get
\beqo
\lefteqn{\phi_i^*(t)=\frac{U_i(t)-\gamma_i(t)\big(-\frac{\theta(t)}{\delta(t)}2K_2(t,Z(t),U(t))+\tilde{\pi}^*(t,Z(t),U(t))\big)}{2K_2(t,Z(t),U(t))}}\\
&=&\frac{\gamma_i(t)}{\delta(t)}\theta(t)\\
&&-\frac{\big(U_i(t)-\tilde{\pi}^*(t,Z(t),U(t))\gamma_i(t)\big)\sqrt{|L(t)|^2-|\theta(t)|^2}}{\sqrt{|\tilde{\pi}^*(t,Z(t),U(t))\sigma(t)-Z(t)|^2+\sj|U_i(t)-\tilde{\pi}^*(t,Z(t),U(t))\gamma_i(t)|^2\lambda_i}},
\eeqo
and condition \eqref{arbitragecondition2} implies that $|\phi_i^*(t)|<1, 0\leq t\leq T, \ i=1,...,I$. Hence, $(\phi^*,\psi^*)\in\mathcal{Q}$. Since there exists a unique solution to the BSDE \eqref{bsdeaa}, we deduce that $Y^{\phi^*,\psi^*}(t)=Y(t), \ 0\leq t\leq T$, and we finally conclude
\beqo
\sup_{(\phi, \psi)\in\mathcal{Q}^{\phi,\psi}}Y^{\phi,\psi}(t)=Y^{\phi^*,\psi^*}(t)=Y(t),\quad 0\leq t\leq T.
\eeqo
\cbdu

In Theorem \ref{nogooddealprice} we formulate conditions which guarantee that our optimal price process \eqref{bsde1} is arbitrage-free and we provide its arbitrage-free representation. The first condition \eqref{arbitragecondition1} excludes from considerations trivial cases of dynamics and claims which do not have a regime-switching component. The key condition which implies that the instantaneous Sharpe ratio pricing is arbitrage-free is the second condition \eqref{arbitragecondition2}. We point out that in a general model with jumps the instantaneous Sharpe ratio pricing can lead to arbitrage prices and some conditions have to be introduced to exclude arbitrage prices, see \cite{delong2011}. Such a condition is proposed in \eqref{arbitragecondition2}. We are aware that \eqref{arbitragecondition2} is not optimal, yet we believe that it should be sufficient in most financial applications. One can notice that our arbitrage-free pricing condition \eqref{arbitragecondition2} is satisfied if the stock's Sharpe ratio $\theta$ is not too large (only required if $\gamma\neq 0$), the surplus' Sharpe ratios $L$ is not too large (compared to $\theta$) and transition intensities $\lambda$ are not too small. Those assumptions should be fulfilled in most cases. In particular, let us remark that \cite{lo} estimates monthly Sharpe ratios for different assets in the range of $(0.14,1.26)$, \cite{hardy} estimates the intensity of transition into a "bad" state at $0.5$ and the intensity of transition into a "good" state at $6$, whereas \cite{hamilton} estimates those intensities at $0.5$ and $1$.

\subsection{Monotonicity of the price}

It is clear that a reasonable pricing operator should be monotone with respect to the terminal claim, in the sense that a more severe claim should be valued at a higher price. Moreover, since the process $L$ appearing in \eqref{objectiveprice} is interpreted as a Sharpe ratio, we should also expect that the higher the Sharpe ratio the hedger requires, the higher the price of the claim should be. Such properties of our optimal price process \eqref{bsde1} could be established provided that we could apply a comparison principle for BSDEs. However, it is well known that a comparison principle for BSDEs with jumps does not always hold, see \citet{buckdahn}, \citet{royer}. Consequently, the price process \eqref{bsde1} may not satisfy the monotonic properties in all cases. In the next theorem we give conditions which guarantee that a comparison principle for the BSDE \eqref{bsde1} can be used and our optimal price process fulfills the desirable monotonic properties.

\begin{thm}\label{comparison}
Let $\xi, \ \xi'$ be $\bfi$-measurable claims such that $\be[|\xi|^2]<\infty, \ \be[|\xi'|^2]<\infty$. Assume that (A1)-(A4) hold and
\beq\label{monotonicity0}
\frac{|\sigma(t)|^2+\sum_{j\neq i}|\gamma_j(t)|^2\lambda_j(t)}{|\delta(t)|^2}(|L(t)|^2-|\theta(t)|^2)<\lambda_i(t),\quad 0\leq t\leq T, \ i=1,...,I,\quad \
\eeq
on the set $\{\gamma_i(t)=0, \ \lambda_i(t)>0\}$, and
\beq\label{monotonicity1}
\lefteqn{\frac{|\sigma(t)|^2+\sum_{j\neq i}|\gamma_j(t)|^2\lambda_j(t)}{|\delta(t)|^2}(|L(t)|^2-|\theta(t)|^2)}\nonumber\\
&&+\frac{|\gamma_i(t)|^2\lambda_i(t)}{|\delta(t)|^2}|\theta(t)|^2<\frac{\lambda_i(t)}{2}\quad 0\leq t\leq T, \ i=1,...,I,
\eeq
on the set $\{\gamma_i(t)\neq0, \ \lambda_i(t)>0\}$. Let $Y$ and $Y'$ denote the solutions to the BSDEs \eqref{bsde1} with terminal conditions $\xi$ and $\xi'$ and coefficients $L$ and $L'$. If $\xi\leq \xi'$ and $L(t)\leq L'(t),\ 0\leq t\leq T$, then $Y(t)\leq Y'(t),\ 0\leq t\leq T$.
\end{thm}
\Proof
Let $f$ denote the generator of the BSDE \eqref{bsde1}. Recalling \eqref{varianceoptimum}-\eqref{representationobjective} we can notice that
\beqo
\lefteqn{f(Y(t),Z(t),U_1,(t),...,U_I(t))-f(Y(t),Z(t),U_1'(t),...,U'_I(t))}\\
&=&\sj\frac{\Gamma_i(t)}{(U_i(t)-U'_i(t))\lambda_i(t)}\mathbf{1}\{(U_i(t)-U'_i(t))\lambda_i(t)\neq 0\}(U_i(t)-U'_i(t))\lambda_i(t),
\eeqo
where
\beq\label{gammacomparison}
\lefteqn{\Gamma_i(t)=\frac{\gamma_i(t)\theta(t)}{\delta(t)}(U_i(t)-U'_i(t))\lambda_i(t)}\nonumber\\
&&-\sqrt{|L(t)|^2-|\theta(t)|^2}\sqrt{|\tilde{\pi}_i(t)\sigma(t)-Z(t)|^2+\sum_{j=1}^I|\tilde{\pi}_i(t)\gamma_j(t)-U_{j,i}(t)|^2\lambda_j(t)}\nonumber\\
&&+\sqrt{|L(t)|^2-|\theta(t)|^2}\sqrt{|\tilde{\pi}_{i+1}(t)\sigma(t)-Z(t)|^2+\sum_{j=1}^I|\tilde{\pi}_{i+1}(t)\gamma_j(t)-U_{j,i+1}(t)|^2\lambda_j(t)},\quad \ \quad \
\eeq
and we introduce
\beqo
\lefteqn{\tilde{\pi}_i(t)}\nonumber\\
&=&\frac{Z(t)\sigma(t)+U'_1(t)\gamma_1(t)\lambda_1(t)+...+U'_{i-1}(t)\gamma_{i-1}(t)\lambda_{i-1}(t)+U_i(t)\gamma_i(t)\lambda_i(t)+...+U_I(t)\gamma_I(t)\lambda_I(t)}{|\delta(t)|^2},\\
\lefteqn{U_{j,i}(t)=U_i(t)\textbf{1}\{j\geq i\}+U'_i(t)\textbf{1}\{j< i\}.}
\eeqo
In order to apply a comparison principle for BSDEs with jumps we have to control the coefficients $\frac{\Gamma_i(t)}{(U_i(t)-U'_i(t))\lambda_i(t)}$, see Theorem 2.5 in \citet{royer}. We show that conditions \eqref{monotonicity0}-\eqref{monotonicity1} imply $|\frac{\Gamma_i(t)}{(U_i(t)-U'_i(t))\lambda_i(t)}|<1, \ 0\leq t\leq T, \ i=1,...,I,$ which is a sufficient condition for the application of the comparison principle from \cite{royer}. We use the equivalent condition
\beq\label{comparison1}
\lefteqn{\Big|\frac{\Gamma_i(t)}{(U_i(t)-U'_i(t))\lambda_i(t)}\Big|<1\Longleftrightarrow}\nonumber\\
&&\Big|\frac{\Gamma_i(t)}{(U_i(t)-U'_i(t))\lambda_i(t)}\Big|^2<1,\quad 0\leq t\leq T,\ i=1,...,I.\quad \
\eeq
By \eqref{norminequality} we get
\beq\label{monotonicitycond}
|\Gamma_i(t)|^2&\leq& 2\frac{|\gamma_i(t)|^2}{|\delta(t)|^2}|\theta(t)|^2\big|(U_i(t)-U'_i(t))\lambda_i(t)\big|^2\nonumber\\
&&+2(|L(t)|^2-|\theta(t)|^2)\Big(|\tilde{\pi}_i(t)\sigma(t)-\tilde{\pi}_{i+1}(t)\sigma(t)|^2\nonumber\\
&&\sum_{j=1}^I|\tilde{\pi}_i(t)\gamma_j(t)-\tilde{\pi}_{i+1}(t)\gamma_j(t)-U_{j,i}(t)+U_{j,i+1}(t)|^2\lambda_j(t)\Big)\nonumber\\
&=&2\frac{|\gamma_i(t)|^2}{|\delta(t)|^2}|\theta(t)|^2\big|(U_i(t)-U'_i(t))\lambda_i(t)\big|^2\nonumber\\
&&+2(|L(t)|^2-|\theta(t)|^2)\Big(\Big|\frac{(U_i(t)-U_i'(t))\gamma_i(t)\lambda_i(t)}{|\delta(t)|^2}\Big|^2|\sigma(t)|^2\nonumber\\
&&+\Big|\frac{(U_i(t)-U_i'(t))\gamma_i(t)\lambda_i(t)}{|\delta(t)|^2}\gamma_i(t)-U_i(t)+U'_{i}(t)\Big|^2\lambda_i(t)\nonumber\\
&&\sum_{j=1, j\neq i}^I\Big|\frac{(U_i(t)-U_i'(t))\gamma_i(t)\lambda_i(t)}{|\delta(t)|^2}\gamma_j(t)\Big|^2\lambda_j(t)\Big)
\eeq
Rearranging the terms in \eqref{monotonicitycond}, it is straightforward to deduce that \eqref{comparison1} is implied by our condition \eqref{monotonicity1} (or \eqref{monotonicity0} if $\gamma_i=0$). The comparison principle now follows from Theorem 2.5 in \citet{royer}.
\cbdu

We have proved the comparison principle under \eqref{monotonicity0}-\eqref{monotonicity1}. It is possible to derive different sufficient conditions for the comparison principle by requiring $|\frac{\Gamma_i(t)}{(U_i(t)-U'_i(t))\lambda_i(t)}|<1$ instead of $|\frac{\Gamma_i(t)}{(U_i(t)-U'_i(t))\lambda_i(t)}|^2<1$. However, the conditions presented in \eqref{monotonicity0}-\eqref{monotonicity1} seem to be more concise and easier to check and interpret. One can notice that \eqref{monotonicity0}-\eqref{monotonicity1} are satisfied, and, consequently, the optimal price process \eqref{bsde1} is a monotonic pricing operator in the sense of Theorem \ref{comparison}, if the stock's Sharpe ratio $\theta$ is not too large (only required if $\gamma\neq 0$), the surplus' Sharpe ratios $L$ is not too large (compared to $\theta$) and transition intensities $\lambda$ are not too small. As discussed at the end of the previous section, those assumptions should hold in most cases. Even though \eqref{monotonicity0}-\eqref{monotonicity1} are not optimal they should be sufficient in applications. Finally, let us remark that if impose a stronger but simpler condition: $|\theta(t)|^2<|L(t)|^2<\lambda_i(t)/2, \ 0\leq t\leq T, i=1,..,I$, then \eqref{arbitragecondition2}, \eqref{monotonicity0}-\eqref{monotonicity1} are satisfied and our optimal price process \eqref{bsde1} is arbitrage-free and fulfills the properties of monotonicity with respect to the terminal claim and the Sharpe ratio.

\subsection{The Markov-regime-switching model}

In practical applications we deal with Markovian models. In this section we establish the relation between the solution to the BSDE \eqref{bsde1} and the solution to a partial integro-differential equation. Such a relation allows us to interpret the optimal hedging strategy and provides a method for finding the solution to our BSDE.

\begin{thm}\label{pidetheorem}
Consider the Markov-regime-switching financial model \eqref{bsregimes} with the pay-off $\xi=F(S(T),J(T))$ and the Sharpe ratio $L(t)=L(J(t-))$. Assume that $\be[|\xi|^2]<\infty$, and let (A1)-(A4) hold. If there exists a unique classical solution $V$ with a uniformly bounded derivative $V_s(t,s,i)$ to the system of nonlinear PIDEs
\beq\label{pidesharpe}
\lefteqn{V_t(t,s,i)+V_s(t,s,i)s\mu(i)+\frac{1}{2}V_{ss}(t,s,i)s^2\sigma^2(i)}\nonumber\\
&&+\sum_{j\neq i}\big(V(t,s+s\gamma_j(i),j)-V(t,s,i)-V_s(t,s,i)s\gamma_j(i)\big)\lambda_j(i,s)=V(t,s,i)r(i)\nonumber\\
&&+\frac{V_s(t,s,i)s\sigma^2(i)+\sum_{j\neq i}\big(V(t,s+s\gamma_j(i),j)-V(t,s,i))\gamma_j(i)\lambda_j(i,s)}{\delta(i)}\theta(i)\nonumber\\
&&-\sqrt{L^2(i)-\theta^2(i)}\sqrt{g(V(t,s,i))},\quad (t,s)\in[0,T)\times(0,\infty), i=1,...,I,\nonumber\\
\lefteqn{V(T,s,i)=F(s,i),\quad s\in(0,\infty), i=1,...,I,}
\eeq
where
\beqo
\lefteqn{g(V(t,s,i))=|V_s(t,s,i)|^2s^2\sigma^2(i)+\sum_{j\neq i} |V(t,s+s\gamma_j(i),j)-V(t,s,i)|^2\lambda_j(i,s)}\\
&&-\frac{\big|V_s(t,s,i)s\sigma^2(i)+\sum_{j\neq i}\big(V(t,s+s\gamma_j(i),j)-V(t,s,i))\gamma_j(i)\lambda_j(i,s)\big|^2}{\delta^2(i)},
\eeqo
then the solution to the BSDE \eqref{bsde1} can be characterized as
\beqo
Y(t)&=&V(t,S(t),J(t)),\quad 0\leq t\leq T,\\
Z(t)&=&V_s(t,S(t-),J(t-))S(t-)\sigma(J(t-)),\quad 0\leq t\leq T,\\
U_i(t)&=&\Big(V(t,S(t-)+S(t-)\gamma_i(J(t-)),i)\\
&&-V(t,S(t-),J(t-))\Big)\mathbf{1}\{i\neq J(t-)\}\quad 0\leq t\leq T, \ i=1,...,I.
\eeqo
\end{thm}
\Proof
From the Markov property of the system we can deduce that $Y(t)=V(t,S(t),J(t)), \ 0\leq t\leq T,$ for some measurable function $V$, see Corollary 2.3 and Remark 2.4 in \cite{buckdahn}. Assuming that $V$ is sufficiently smooth, we can apply the It\^{o}'s formula and we get the dynamics
\beq\label{itoformula}
\lefteqn{dV(t,S(t),J(t))=V_t(t,S(t-),J(t-))dt+V_s(t,S(t-),J(t-))S(t-)\mu(J(t-))dt}\nonumber\\
&&+V_{s}(t,S(t-),J(t-))S(t-)\sigma(J(t-))dW(t)\nonumber\\
&&+\frac{1}{2}V_{ss}(t,S(t-),J(t-))S^2(t-)\sigma^2(J(t-))\nonumber\\
&&+\sum_{j\neq J(t-)}\Big(V(t,S(t-)+S(t-)\gamma_j(J(t-)),j)-V(t,S(t-),J(t-))\Big)d\tilde{N}_j(t)\nonumber\\
&&+\sum_{j\neq J(t-)}\Big(V(t,S(t-)+S(t-)\gamma_j(J(t-)),j)-V(t,S(t-),J(t-))\nonumber\\
&&\quad -V_s(t,S(t-),J(t-))S(t-)\gamma_j(J(t-))\Big)\lambda_j(J(t-),S(t-))dt.
\eeq
The result now follows by comparing the terms in \eqref{itoformula} and \eqref{bsde1}.
\cbdu

Since the process $Y$ models the price of the claim, the function $V$ determines the value of the claim given the current value of the underlying risk factors. It is straightforward to notice that the optimal hedging strategy \eqref{strategy1} consists of two terms. In the view of Theorem \ref{pidetheorem}, the first term is based on the changes in the price of the claim which result from continuous changes in the stock value (the interpretation of the control process $Z$) and a discontinuous change in the stock value induced by a transition of the economy into a new state (the interpretation of the control processes $U$). Hence, the first term of the optimal hedging strategy \eqref{valuenonregular} is a delta-hedging strategy. The second term of the optimal hedging strategy \eqref{strategy1} can be seen as a correction factor for the delta-hedging strategy. Recalling the interpretation of the strategy \eqref{varianceoptimum}, we can deduce that the correction term arises since the hedger optimizes the mean-variance risk measure of the surplus instead of minimizing the variance of the surplus. The correction term reflects the use of the expected profit in the hedging objective, it leads to a higher expected profit of the surplus but also to a higher variance.

Finally, we remark that it would be difficult to establish the existence of a unique classical (or viscosity) solution $V$ to the system of PIDEs \eqref{pidesharpe}. Hence, we believe that our approach based on BSDEs is mathematically more tractable.

\section{The insurance application}
As discussed in Introduction, regime-switching models are very useful for long-term insurance and financial risk management. In this section we consider the non-Markovian regime-switching financial model \eqref{bondregimes}-\eqref{stockregimes} and a mortality-contingent claim with a financial component (an equity-linked insurance claim). We use the instantaneous mean-variance risk measure \eqref{objectivestrategy} and the instantaneous Sharpe ratio \eqref{objectiveprice} to price and hedge the financial risk and the mortality risk.

Let $\tau$ denote the future lifetime of a policyholder. The random variable $\tau$ is exponentially distributed
\beq\label{insurance}
\bp(\tau>t)=e^{-\int_0^tm(s)ds},\quad 0\leq t\leq T,
\eeq
with mortality intensity $m$. Let us set $M(t)=\mathbf{1}\{\tau\geq t\}$ and $\tilde{M}(t)=M(t)-\int_0^t(1-M(s-))m(s)ds, \ 0\leq t\leq T$. As always, we assume that the insurance risk is independent of the financial market. Recalling the results of the previous sections, we can deduce that the optimal price process $Y$ of the claim $\xi$ contingent on the financial risk \eqref{bondregimes}-\eqref{stockregimes} and the insurance risk \eqref{insurance} solves the following BSDE
\beq\label{optimalpriceprocessfinancialmortalityrisks}
\lefteqn{Y(t)=\xi+\int_t^T\Big(-Y(s)r(s)-\frac{Z(s)\sigma(s)+\sj U_i(s)\gamma_i(s)\lambda_i(s)}{\delta(s)}\theta(s)}\nonumber\\
&&+\sqrt{|L(s)|^2-|\theta(s)|^2}\nonumber\\
&&\cdot\sqrt{|Z(s)|^2+\sj |U_i(s)|^2\lambda_i(s)-\frac{|Z(s)\sigma(s)+\sj U_i(s)\gamma_i(s)\lambda_i(s)|^2}{|\delta(s)|^2}+|U_m(s)|^2m(s)} \ \Big)ds \ \quad \ \nonumber\\
&&-\int_t^TZ(s)dW(s)-\int_t^T\sj U_i(s)d\tilde{N}_i(s)-\int_t^TU_m(s)d\tilde{M}(s),\quad 0\leq t\leq T,
\eeq
and the optimal hedging strategy takes the form
\beqo
\lefteqn{\pi^*(t)=\frac{Z(t)\sigma(t)+\sj U_i(t)\gamma_i(t)\lambda_i(t)}{|\delta(t)|^2}}\nonumber\\
&&+\frac{\theta(t)}{\delta(t)\sqrt{|L(t)|^2-|\theta(t)|^2}}\nonumber\\
&&\cdot\sqrt{|Z(t)|^2+\sj |U_i(t)|^2\lambda_i(t)-\frac{|Z(t)\sigma(t)+\sj U_i(t)\gamma_i(t)\lambda_i(t)|^2}{|\delta(t)|^2}+|U_m(t)|^2m(t)},\quad 0\leq t\leq T.
\eeqo
Hence, the results derived in Sections 4-5 can also be applied to solve pricing and hedging problems in combined financial and insurance models. We remark that under some assumptions the optimal price process \eqref{optimalpriceprocessfinancialmortalityrisks} has an arbitrage-free representation and the mortality risk is priced with a nonzero risk premium.

\section{Numerical example}

In this last section we present some numerical results. We consider the Markov-regime-switching model \eqref{bsregimes} with 2 states of economy and the parameters which are specified in Table 1. In particular, the stock's Sharpe ratios are equal to $\theta(1)=0.24$ and $\theta(2)=0.04$.

\begin{table}[h]
\begin{center}
  \caption{The parameters of the Markov-regime-switching model.}
  \begin{tabular}[t]{|r|r|r|r|r|r|}
    \hline
    State $i$ & $r(i)$ & $\mu(i)$ & $\sigma(i)$ & $\gamma_.(i)$ & $\lambda_.(i)$\\ \hline
    1 & 0.03 & 0.07 & 0.1 & -0.1 & 2\\ \hline
    2 &  0.01 & 0,02 & 0.25 & 0.05 & 5\\ \hline
  \end{tabular}
\end{center}
\end{table}

\begin{table}[h]
\begin{center}
  \caption{The prices of the call options with strike $Q$ and the Sharpe ratios $L(1)=0.4, L(2)=0.2$ in the Markov-regime-switching model.}
  \begin{tabular}[t]{|r|r|r|r|r|r|}
    \hline
    The strike & $Q=80$ & $Q=90$ & $Q=100$ & $Q=110$ & $Q=120$\\ \hline
    The price & 24.561 & 16.677 & 10.381 & 5.857 & 2.928\\ \hline
  \end{tabular}
\end{center}
\end{table}

\begin{table}[h]
\begin{center}
  \caption{The prices of the call options with strike $Q$ in the complete Black-Scholes model with parameters $(r, \sigma)$.}
  \begin{tabular}[t]{|r|r|r|r|r|r|}
    \hline
    The strike& $Q=80$ & $Q=90$ & $Q=100$ & $Q=110$ & $Q=120$\\ \hline
    The BS price for $(0.03,0.1)$ & 22.381 & 13.038 & 5.581 & 1.596 & 0.299\\ \hline
    The BS price for $(0.01,0.25)$ &  22.891 & 15.830 & 10.405 & 6.532 & 3.948\\ \hline
  \end{tabular}
\end{center}
\end{table}

\begin{table}[h]
\begin{center}
  \caption{The prices of the call options with the strike $Q=100$ and Sharpe ratios $(L(1),L(2))$ in the Markov-regime-switching model.}
  \begin{tabular}[t]{|r|r|r|r|r|r|}
    \hline
    The Sharpe ratios & $L(1)=0.24$ & $L(1)=0.3$& $L(1)=0.5$& $L(1)=0.7$&$L(1)=1.2$\\
    &$L(2)=0.04$ & $L(2)=0.1$ & $L(2)=0.3$ & $L(2)=0.5$ & $L(2)=1.2$\\ \hline
    The price & 9.827 & 10.082 & 10.660 & 11.226 & 13.168\\ \hline
  \end{tabular}
\end{center}
\end{table}

We are interested in pricing 1-year call options with various strikes $Q$. The initial price of the derivative is determined by the solution $Y(0)$ to the BSDE \eqref{bsde1}. The BSDE has to be solved numerically. We apply discrete-time approximation and Least Squares Monte Carlo. In our example the solution to the BSDE \eqref{bsde1} can be derived by using the backward recursion
\beqo
\lefteqn{Y(1)=(S(T)-Q)^+,}\\
\lefteqn{Z_i(t_k)=\frac{1}{h}\be\big[Y(t_{k+1})(W(t_{k+1})-W(t_k))|S(t_k)=s,J(t_k)=i\big],\quad i=1,2,}\\
\lefteqn{U_1(t_k)=\frac{1}{\lambda_1(2)h}\be\big[Y(t_{k+1})(\tilde{N}_1(t_{k+1})-\tilde{N}_1(t_k))|S(t_k)=s,J(t_k)=2\big],}\\
\lefteqn{U_2(t_k)=\frac{1}{\lambda_2(1)h}\be\big[Y(t_{k+1})(\tilde{N}_2(t_{k+1})-\tilde{N}_2(t_k))|S(t_k)=s,J(t_k)=1\big],}\\
\lefteqn{Y_i(t_k)=\frac{1}{1+r(i)h}\Big\{\be[Y(t_{k+1})|S(t_k)=s,J(t_k)=i]}\\
&&-\Big(\frac{Z_i(t_k)\sigma(i)+ U_j(t_k)\gamma_j(i)\lambda_j(i)\mathbf{1}\{j\neq i\}}{\delta(i)}\theta(i)\nonumber\\
&&-\sqrt{|L(i)|^2-|\theta(i)|^2}\nonumber\\
&&\cdot\sqrt{|Z_i(t_k)|^2+|U_j(t_k)|^2\lambda_j(i)\mathbf{1}\{j\neq i\}-\frac{|Z_i(t_k)\sigma(i)+U_j(t_k)\gamma_j(i)\lambda_j(i)\mathbf{1}\{j\neq i\}|^2}{|\delta(i)|^2}} \ \Big)h\Big\},\quad i=1,2,\quad \
\eeqo
where $0=t_0<t_1<...<t_{k-1}<t_k=1$ and $h$ is a time-discretization step, and by approximating the processes $Y,Z,U$ with regression functions, see \cite{elie}.

The prices of the call options with various strikes for the Sharpe ratios $L(1)=0.4, L(2)=0.2$ are given in Table 2. Monotonicity of the price with respect to the strike can be observed. In Table 3 we also present the prices of the call options in two classical Black-Scholes model with the parameters $(r(i),\sigma(i))$ determined by the state 1 and 2. We might have expected that the price in the regime-switching model should be between the prices in the Black-Scholes models. However, under our pricing method the hedger specifies his expected profit reflected by the Sharpe ratio $L$ which increases the price. Hence, our price can be between the Black-Scholes prices or above the higher price. The relation between our prices in the regime-switching model and the prices in the Black-Scholes models can be observed by comparing the results in Table 2 and 3. In Table 4 we find the prices of the call options with the strike $Q=100$ for various Sharpe ratios. Monotonicity of the price with respect to the hedger's Sharpe ratio can be observed. We remark that the pair $(0.24,0.04)$ is the lowest Sharpe ratio and $(1.2,1.2)$ is the highest Sharpe ratio (assuming that $L(1)\geq L(2)$) which can be used under the assumptions of Theorems \ref{pricehedge}, \ref{nogooddealprice}, \ref{comparison}. Consequently, in our example for all reasonable values of the hedger's Sharpe ratios the arbitrage-free condition \eqref{arbitragecondition2} and the monotonicity condition \eqref{monotonicity1} are fulfilled.

\section{Conclusion}

We have studied hedging and pricing of unattainable contingent claims in a non-Markovian regime-switching financial model. We have derived the hedging strategy which minimizes the instantaneous mean-variance risk of the hedger's surplus and the price under which the instantaneous Sharpe ratio of the hedger's surplus equals a predefined target. The optimal hedging strategy and the optimal price process have been characterized with a unique solution to a nonlinear, Lipschitz BSDE with jumps. We have discussed key properties of the price and the hedging strategy. Our solution can be applied in practice to value complex, long-term financial and insurance liabilities.


\begin{thebibliography}{99}

\bibitem[Barles et. al.(1997)]{buckdahn}
Barles, G., Buckdahn, R., Pardoux, E. (1997) Backward stochastic differential equations and integral-partial differential equations. \emph{Stochastic and Stochastic Reports} 60, 57-83.
\bibitem[Bayraktar and Young(2008)]{youngfin}
Bayraktar, E., Young, V. (2008) Pricing options in incomplete equity markets via the instantaneous Sharpe ratio. \emph{Annals of Finance} 4, 399-429.
\bibitem[Bayraktar et. al.(2009)]{youngannuity}
Bayraktar, E., Milevsky, M., Promislow, S., Young, V. (2009) Valuation of mortality risk via the instantaneous Sharpe ratio: applications to life annuities. \emph{Journal of Economic Dynamic and Control} 33, 676-691.
\bibitem[Becherer(2006)]{Bech}
Becherer, D. (2006) Bounded solutions to BSDE's with jumps for
utility optimization and indifference hedging. \emph{The Annals of Applied
Probability} 16, 2027-2054.
\bibitem[Becherer(2009)]{Bech2}
Becherer, D. (2009) From bounds on optimal growth towards a theory of good-deal hedging. In \emph{Advanced Financial Modelling} (ed. H. Albecher, W. Runggaldier, W. Schachermayer), pp. 27-52, de Gruyter.
\bibitem[Bj\"{o}rk and Slinko(2006)]{bjork}
Bj\"{o}rk, T. Slinko, I. (2006) Towards a general theory of good deal bounds. \emph{Review of Finance} 10, 221-260.
\bibitem[Bouchard and Elie(2008)]{elie}
Bouchard, B., Elie, R. (2008) Discrete time approximation of decoupled forward backward SDE with jumps. \emph{Stochastic Processes and their Applications} 118, 53-75.
\bibitem[Cochrane and Sa\'{a}-Requejo(2000)]{cr}
Cochrane, J., Sa\'{a}-Requejo J. (2000) Beyond arbitrage: good-deal asset price bounds in incomplete markets. \emph{Journal of Political Economy} 1008, 79-119.
\bibitem[Cr\'{e}pey(2011)]{crepeymodel}
Cr\'{e}pey, S. (2011) About the pricing equation in finance. In: Carmona, R.A., \c{C}inlar, E., Ekeland, I., Jouini, E., Scheinkman, J.A., Touzi, N. (eds.) \emph{Paris-Princeton Lectures in Mathematical Finance 2010}, pp. 63-203. Springer.
\bibitem[Delong(2011)]{delong2011}
Delong, {\L}. (2011) No-good-deal, local mean-variance and ambiguity risk pricing and hedging for an insurance payment process. \emph{ASTIN Bulletin} 42, 203-232.
\bibitem[Donnelly(2010)]{donely}
Donnelly, C. (2011) Good-deal bounds in a regime switching market. \emph{Applied Mathematics and Optimization} 18, 491-515.
\bibitem[Donnelly and Heunis(2012)]{donely2}
Donnelly, C., Heunis, A.J. (2012) Quadratic risk minimization in a regime-switching with portfolio constraints, \emph{SIAM Journal of Control and Optimization} 50, 2431-2461.
\bibitem[Elliott et. al.(2011)]{elliotstochastic}
Elliot, R.J., Siu, T.K., Chan, L.L., Lau, J.W. (2007) Pricing options under a generalized Markov-modulated jump-diffusion model. \emph{Stochastic Analysis and Applications} 35, 694-713.
\bibitem[Elliott et. al.(2010)]{elliotannals}
Elliot, R.J., Siu, T.K. (2010) On risk minimizing portfolios under a Markovian regime-switching Black-Scholes economy. \emph{Annals of Operations Research} 176, 271-291.
\bibitem[Elliott et. al.(2011)]{elliot}
Elliot, R.J., Siu, T.K., Badescu, A. (2011) On pricing and hedging options in regime-switching models with feedback effect. \emph{Journal of Economic Dynamics and Control} 35, 694-713.
\bibitem[El Karoui et al.(1997)]{karouipengbsde}
El Karoui, N., Peng, S., Quenez, M.C. (1997) Backward stochastic
differential equations in finance.\emph{ Mathematical Finance} 7, 1--71.
\bibitem[Hamilton(1989)]{hamilton}
Hamilton, J.D. (1989) A new approach to the economic analysis of nonstationary time series and the business cycles. \emph{Econometrica} 57, 357-384.
\bibitem[Hardy(2001)]{hardy}
Hardy, M. (2001) A regime-switching model of long-term stock returns. \emph{North American Actuarial Journal} 52, 41-53.
\bibitem[He et. al.(1992)]{Y}
He, S., Wang, J., Yan, J. (1992) \emph{Semimartingale Theory and
Stochastic Calculus}. CRC Press.
\bibitem[Jacod and Shiryaev(2003)]{jacod}
Jacod, J., Shiryaev, A.N. (2003) \emph{Limit Theorems for Stochastic Processes}. Springer.
\bibitem[Leitner(2007)]{leitner}
Leitner, J. (2007) Pricing and hedging with globally and instantaneously vanishing risk. \emph{Statistics and Decisions} 25, 311-332.
\bibitem[Lo(2002)]{lo}
Lo, A. W. (2002) The statistics of Sharpe ratios. \emph{Financial Analysis Journal} 58, 36-52.
\bibitem[Mamon and Elliot(2007)]{eliotbook}
Mamon, R.S, Elliot, R.J. (2007) \emph{Hidden Markov Models in Finance}. Springer.
\bibitem[Pelsser(2011)]{pelsser}
Pelsser, A. (2011) Pricing in incomplete markets. Research paper, http:/ /papers.ssrn.com/sol3/papers.cfm?abstract\_id=1855565.
\bibitem[Protter(2004)]{Protter}
Protter, P.(2004) \emph{Stochastic Integration and Differential
Equations}. Springer.
\bibitem[Royer(2006)]{royer}
Royer, M. (2006) Backward stochastic differential equations with jumps and related non-linear expectations. \emph{Stochastic Processes and their Applications} 116, 1358-1376.
\bibitem[Siu et. al.(2008)]{siu}
Siu, T.K., Erlein, C., Mamon, R.S.(2008) The pricing of credit default swaps under a Markov-modulated Merton's structural model. \emph{North American Actuarial Journal} 12, 19-46.
\bibitem[Siu(2012)]{siuime}
Siu, T.K. (2012) A game theoretic approach to option valuation under Markovian regime-switching models. \emph{Insurance: Mathematics and Economics} 42, 1146-1158.
\bibitem[Wu and Li(2012)]{wu}
Wu, H., Li, Z. (2012) Multi-period mean–variance portfolio selection with regime switching and a stochastic cash flow. \emph{Insurance: Mathematics and Economics} 50, 371-384.
\bibitem[Young(2008)]{younglife}
Young, V. (2008) Pricing life insurance under stochastic mortality via the instantaneous Sharpe ratio. \emph{Insurance: Mathematics and Economics} 42, 691-703.

\end{thebibliography}
\end{document}